\documentclass[]{spie}  
\usepackage[]{graphicx,url}

\title{Modelling electron distributions within ESA's Gaia satellite CCD pixels to mitigate radiation damage} 

\author{G. M. Seabroke\supit{*a}, A. D. Holland\supit{a}, D. Burt\supit{b}, M. S. Robbins\supit{b}
\skiplinehalf
\supit{a}e2v Centre for Electronic Imaging, Planetary \& Space Sciences Research Institute, \\The Open University, Milton Keynes, UK; \\
\supit{b}e2v technologies, Chelmsford, UK
}

\authorinfo{*g.m.seabroke@open.ac.uk}

\begin{document} 
  \maketitle 

\begin{abstract}
The Gaia satellite is a high-precision astrometry, photometry and spectroscopic ESA cornerstone mission, currently scheduled for launch in 2012. Its primary science drivers are the composition, formation and evolution of the Galaxy. Gaia will achieve its unprecedented positional accuracy requirements with detailed calibration and correction for radiation damage.  At L2, protons cause displacement damage in the silicon of CCDs.  The resulting traps capture and emit electrons from passing charge packets in the CCD pixel, distorting the image PSF and biasing its centroid.  Microscopic models of Gaia's CCDs are being developed to simulate this effect. The key to calculating the probability of an electron being captured by a trap is the 3D electron density within each CCD pixel. However, this has not been physically modelled for the Gaia CCD pixels. In Seabroke, Holland \& Cropper (2008), the first paper of this series, we motivated the need for such specialised 3D device modelling and outlined how its future results will fit into Gaia's overall radiation calibration strategy. In this paper, the second of the series, we present our first results using Silvaco's physics-based, engineering software: the ATLAS device simulation framework.  Inputting a doping profile, pixel geometry and materials into ATLAS and comparing the results to other simulations reveals that ATLAS has a free parameter, fixed oxide charge, that needs to be calibrated. ATLAS is successfully benchmarked against other simulations and measurements of a test device, identifying how to use it to model Gaia pixels and highlighting the effect of different doping approximations. 
\end{abstract}



\section{INTRODUCTION}
\label{sec:intro}  

The primary science drivers of Gaia, the ESA cornerstone mission scheduled for launch in 2012, are the composition, formation and evolution of the Galaxy.  Gaia will map the positions and kinematics of $10^{9}$ Galactic stars complete to $V = 20$ mag with unprecedented precision.  However, Gaia's required centroiding accuracy per observation, on the order of one-thousandth to one-tenth of a pixel dependent on magnitude, is already challenging the calibration of today's state-of-the-art CCDs.  Gaia's CCDs were specifically designed for Gaia and are currently being manufactured by e2v technologies (UK).  There are many additional calibration challenges facing Gaia\cite{kohley2009} (see Ref. 1 for a more detailed discussion of Gaia's operational aspects and tests of Flight Model CCDs with respect to calibration needs).  This paper focuses on the calibration challenge of needing to correct for radiation damage.  

Operating at L2, the satellite is expected to experience an end-of-mission (after five years) particle fluence of $\approx$$5 \times 10^{9}$ (mainly solar) protons cm$^{-2}$ (10 MeV equivalence).  In the silicon (Si) of the 106 Gaia CCDs, this will cause displacement damage in the form of electron traps.  The CCDs are operated in Time Delay and Integrate (TDI) mode because the satellite will continuously spin (and precess).  Depending on trap occupancy level determined by illumination history, traps may capture electrons from charge packets in pixels in the leading edge of image point spread functions (PSFs).  Slow-emission (compared to the TDI period) traps can release their captured electrons after the image has passed.    This charge loss from the image leads to lower image signal-to-noise.  Periodic charge injection can keep the slow traps full but cannot fill the fast-emission traps\cite{seabroke2008} (see Ref. 2 for more details of the trap types).  Fast traps can release their captured electrons into charge packets in pixels in the PSF trailing edge.  This will distort the PSF shape such that it no longer matches the fitting function used to centroid the image.  The distorted PSF leads to a shifted centroid.  This effect has been measured to be up to on the order of one-tenth of a pixel, which, depending on the brightness of the object, can be greater than the required centroiding accuracy.%

Gaia's approach is to have a radiation damage correction model that can produce a range of distorted fitting functions caused by a range of damage levels.  The current approach does not fit a single distorted PSF to the data but instead must produce a damaged image at each iteration of the image parameter determination process. This is because the damaged profile depends on the sub-pixel location of the source and so is tied closely to the `true' image parameters.  The most accurate radiation damage model is assumed to be a microscopic implementation of the Shockley-Read-Hall theory\cite{shockley1952,hall1952} that models individual traps and charge packets.  Monte Carlo techniques are required to model the stochastic probability of traps capturing and releasing electrons and to average over a range of trap positions within a pixel.  The probability of an empty trap capturing an electron depends on the electron velocity (determined by known fundamental constants), trap cross-section (approximately known for some traps but often left as a free fitting parameter) and electron density in the vicinity of the trap (unknown but charge packets within pixels can be physically modelled by simultaneously solving for electrostatic potential and charge density using the Poisson and charge continuity equations respectively).  Ongoing radiation test campaigns irradiate Gaia CCDs providing limited parameter space test data from which the effects of radiation damage on Gaia data have been derived and with which microscopic models can be verified. Subsequently, microscopic models can provide full parameter space synthetic data to test Gaia's data processing pipeline.

However, microscopic models are currently too slow to be included in Gaia's data processing pipeline due to the extremely high data aquisition rate.  Instead, much faster and consequently simpler macroscopic (e.g. column integrated) models are required to conduct the radiation damage correction to the data.  These have developed from microscopic models and require microscopic models (as well as test data) to rigorously test them.  One of the proposed macroscopic models\cite{short2009} is based on how the volume of a charge packet grows with its number of constituent electrons.  Hence, both microscopic and macroscopic models require knowledge of electron distribution as a function of position within a pixel and as a function of the number of constituent electrons as inputs.  Therefore the modelling of these inputs for Gaia's e2v CCD91-72 pixels using commercially available, physics-based, three dimensional (3D) device simulation software by Silvaco was proposed\cite{seabroke2008}.  

This paper, the second in the series, presents the first results: benchmarking the Silvaco software.  Because it is much easier to physically measure electrostatic potential ($\phi$) than electron distributions, benchmarking of this type generally consists of simulating a device's $\phi$ distribution and comparing it to that measured in the device.  As $\phi$ determines the electron distribution, if the simulated $\phi$ agrees with reality, this suggests the simulated electron distribution will also agree with reality.  Silvaco is benchmarked against other simulation software and physical measurements.  Section \ref{sec:process} introduces e2v's process parameters, which guide the benchmarking exercise.  The results are presented in Sec.~\ref{sec:bench}, with conclusions and future work in Sec.~\ref{sec:conc}.

\section{e2v process parameters}
\label{sec:process}

\begin{figure}[htbp]
\begin{center}
 \includegraphics[width=0.49\textwidth]{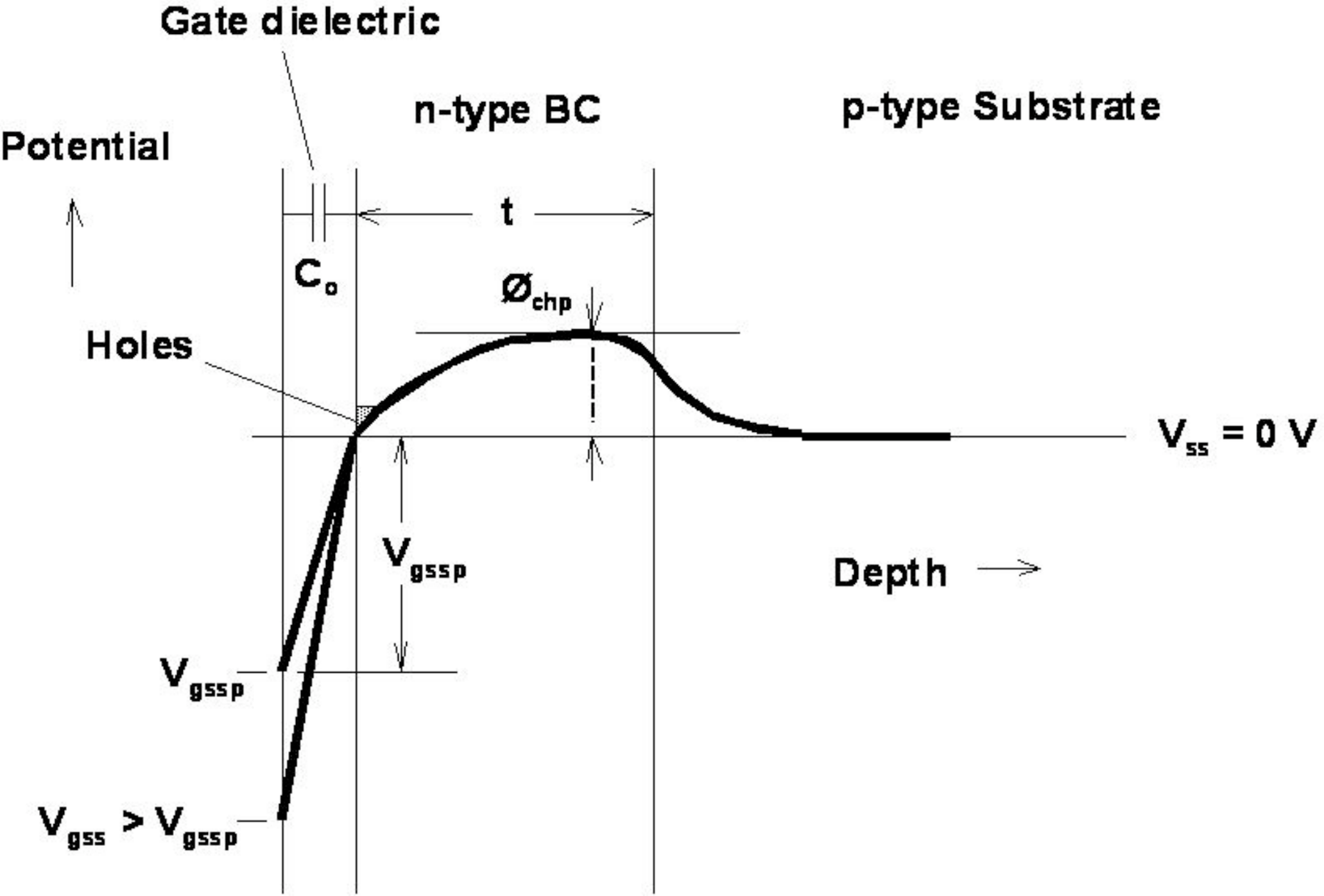}
  \includegraphics[width=0.49\textwidth]{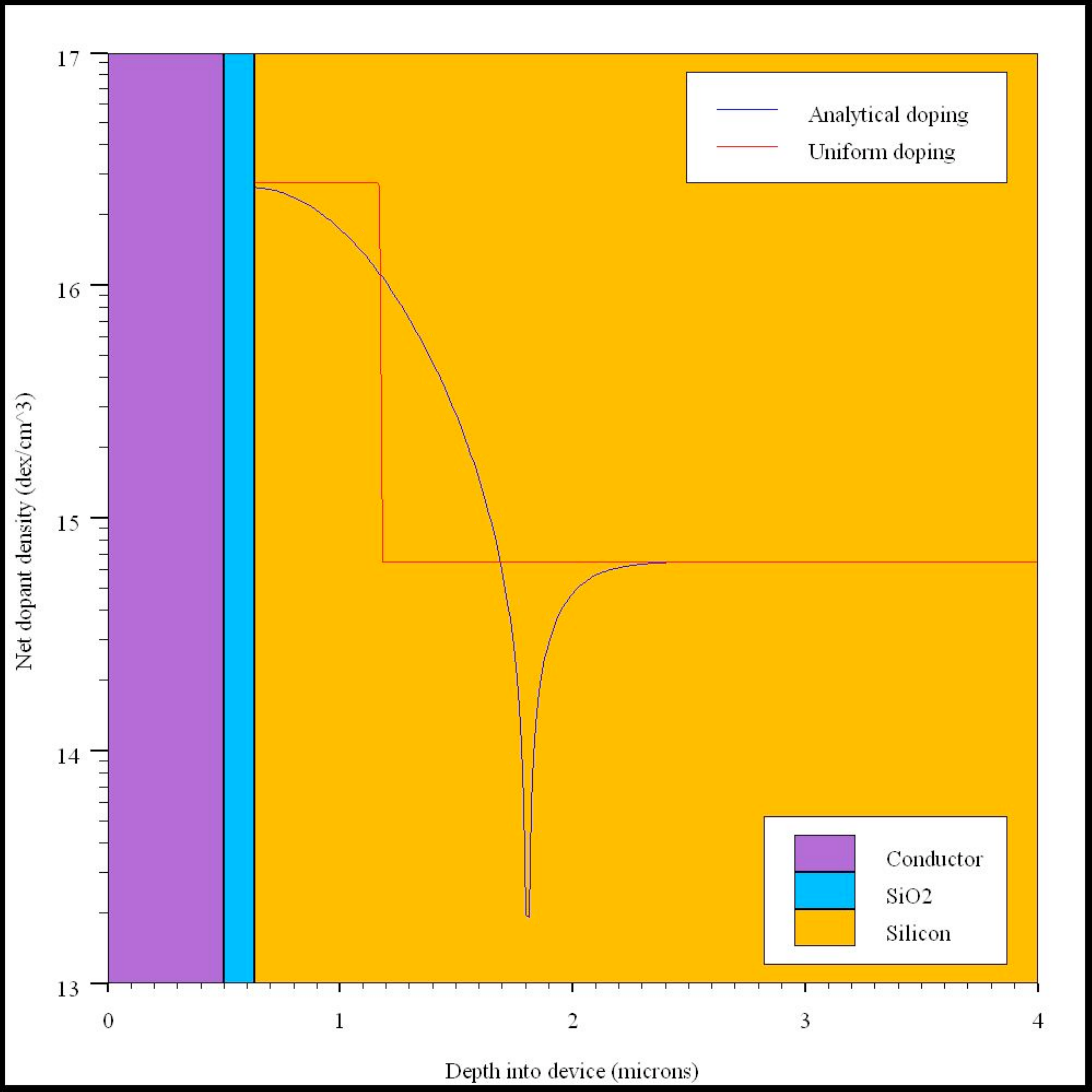}
 \end{center}
\caption{{\it Left}: Schematic $\phi$ profile through an insulating oxide layer (gate dielectric), buried channel (BC) and substrate when a CCD pixel is in Inverted Mode Operation (IMO).  The line between the n-type BC and p-type substrate is the p-n junction.  {\it Right}:   Silvaco TonyPlot comparing realistic and step (box) doping profiles: net dopant density as a function of depth into the device.  The background is colour-coded according to the material structure of the ATLAS models as a function of depth: conductor (polysilicon electrodes, depth 0-0.5 $\mu$m), SiO$_{2}$ (insulator, depth 0.5-0.63 $\mu$m) and silicon (semiconductor, depth 0.63-16 $\mu$m).  Because the doping density is absolute, the minimum point on the realistic doping profile is the p-n junction, separating p-type ions to the right and n-type ions to the left.}
\label{fig:pot_profile}
\end{figure}

\noindent If the voltage difference between an electrode or gate (g) and the substrate (ss) is $V_{\textrm{gss}}$ and the voltage on the underlying p-type substrate is $V_{\textrm{ss}}$, then the $\phi$ distribution through the buried channel (BC) part of the structure is as shown in Fig.~\ref{fig:pot_profile}. The maximum $\phi$ in the (buried) channel is designated $\phi_{\textrm{ch}}$. In the case of no signal charge being stored:
$\phi_{\textrm{ch}} \approx SV_{\textrm{gss}} + \phi_{\textrm{ch0}}$,
where $\phi_{\textrm{ch0}}$ is the value of $\phi_{\textrm{ch}}$ for $V_{\textrm{gss}} = 0$~volts (V). $\phi_{\textrm{ch}}$ is termed the ``channel parameter" and essentially relates the voltage in the underlying channel to that applied to the electrode. The actual value depends on the oxide capacitance per unit area, the doping distribution in the channel, the substrate doping concentrations and any fixed charge in the oxide. The factor $S$ also depends on these quantities, being unity for lightly doped substrates and reducing slightly as the doping increases. 

If the voltage on the electrode is taken increasingly negative of the substrate, there comes a point when the voltage at the insulator-semiconductor (silicon dioxide-Si) interface becomes equal to that of the substrate, which allows electron holes to flow in from the substrate and accumulate at the interface. The resulting $\phi$ distribution is as shown in the left schematic of Fig.~\ref{fig:pot_profile}. If the electrode voltage is taken further negative, more holes accumulate and $\phi_{\textrm{ch}}$ does not change and is said to be ``pinned". This maximum pinned $\phi$ value is designated $\phi_{\textrm{chp}}$ and the gate voltage at the onset of pinning is designated $V_{\textrm{gssp}}$.  The quantities $\phi_{\textrm{ch0}}$, $\phi_{\textrm{chp}}$ and $V_{\textrm{gssp}}$ can all be measured using test transistors included around the periphery of a device and values thereby provide useful data for process control (and benchmarking device simulations).

\section{BENCHMARKING} 
\label{sec:bench}


\begin{figure}[htbp]
\begin{center}
 \includegraphics[width=0.49\textwidth]{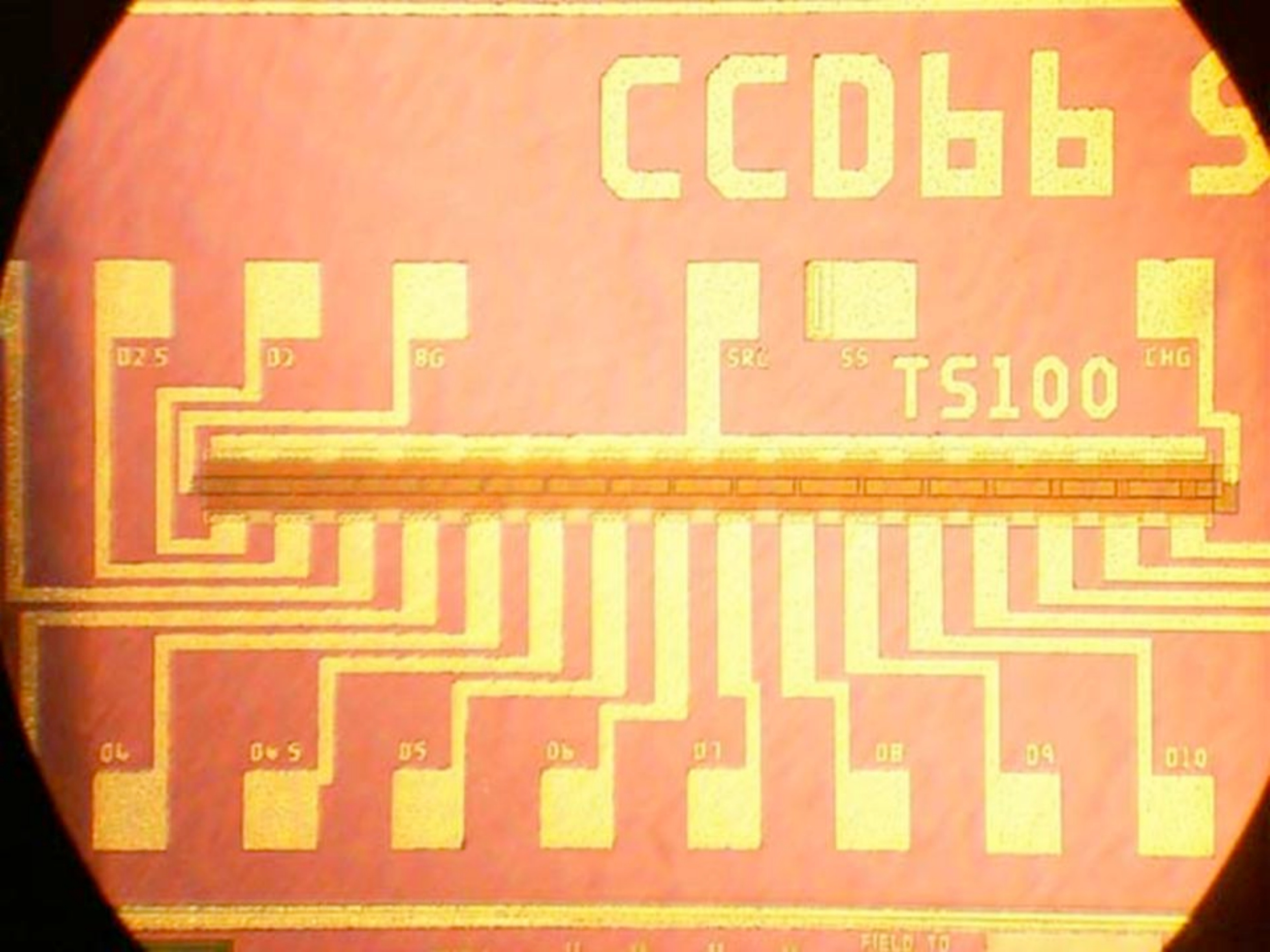}
  \includegraphics[width=0.49\textwidth]{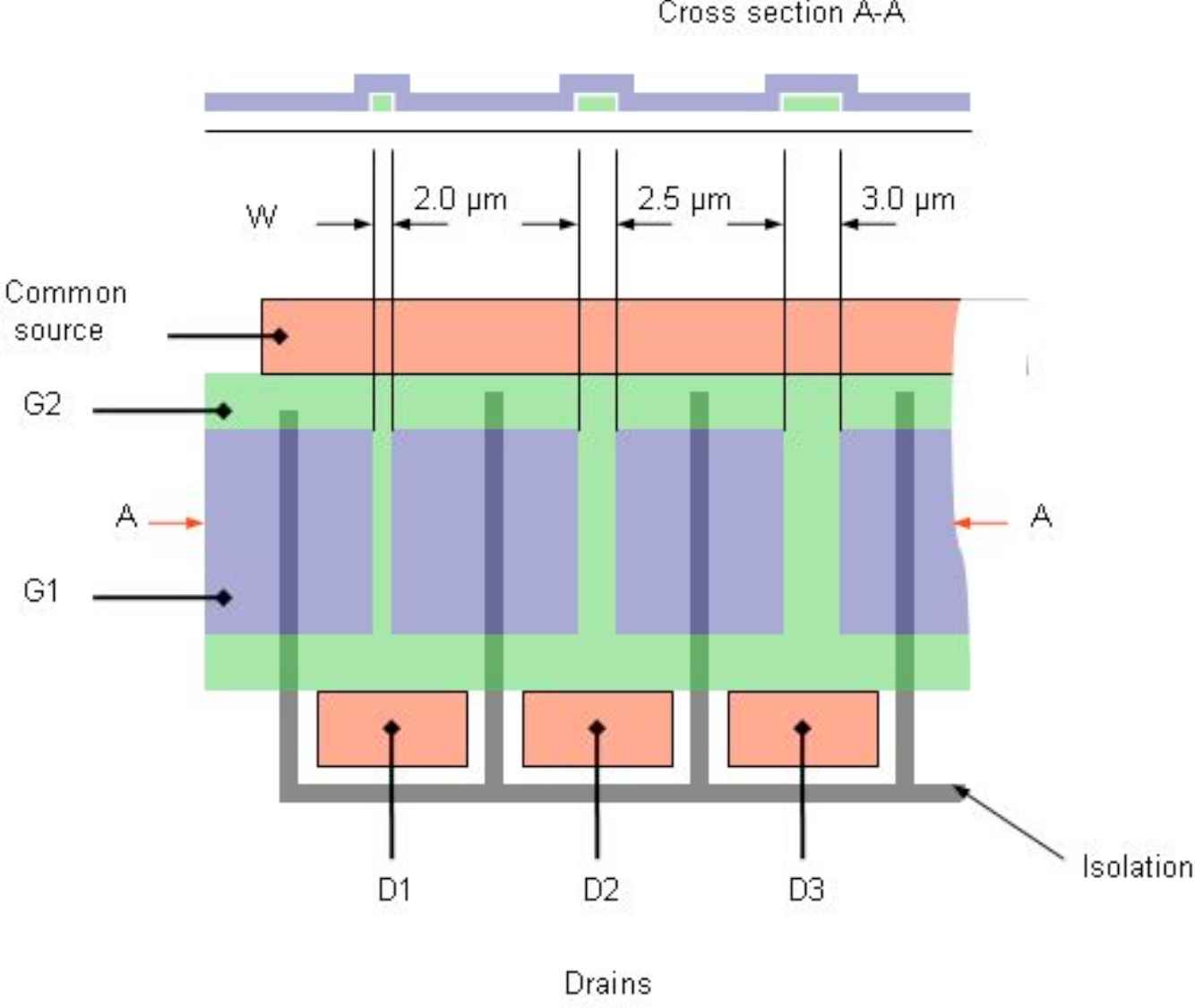}
 \end{center}
 \caption{{\it Left}: Microscopic image of TS100 on a CCD66 wafer.  The connection to the common source is labelled \texttt{SRC} (just above centre of the image) and common source itself runs the entire length of the transistor.  \texttt{SS} labels the connection to the underlying substrate made on the front surface of the device.  The overlying gate (labelled \texttt{BG} for bias gate and annotated G1 in the schematic) is visible along the entire length of the transistor.  The increasing width of the underlying gate (labelled \texttt{CHG} for channel gate and annotated G2 in the schematic) from left to right is only apparent due to the changing topography of G1.  The widths of the narrowest gates are barely visible at the left end of the transistor but the individual drain connections below the transistor indicate the gate positions.  Each individual drain connection is labelled with the width of the gate.   The drain connection to the narrowest gate is labelled \texttt{02} (2 $\mu$m) just to the left of \texttt{BG} above the transistor and connected to the drain below the transistor.  The drain connection to the next narrowest gate is labelled \texttt{02.5} just to the left of \texttt{02} above the transistor and connected to the drain below the transistor to the right of the \texttt{02} drain.  The labelling of the drain connections to gates \texttt{03} and \texttt{03.5} is off the left edge of the image above and below the transistor respectively.  Drain connections to gates \texttt{04}, \texttt{04.5}, \texttt{05}, \texttt{06}, \texttt{07}, \texttt{08}, \texttt{09} and \texttt{10} are labelled along the bottom of the image.  The labelling of the drain connections to gates \texttt{12}, \texttt{14}, \texttt{16} and \texttt{20} is off the right edge of the image.  {\it Right}: Schematic of TS100, showing the narrowest 3 gates:  2, 2.5 and 3 $\mu$m.  W is the width of gate G2 across cross-section A-A.  A vertical view through cross-section A-A is at the top of the schematic where gate G1 overlays gate G2.}
\label{fig:ts100}
\end{figure}


\noindent e2v fabricated a test structure,  a custom field effect transistor (FET), called TS100 from a CCD66 (non-Gaia) wafer.  Figure~\ref{fig:ts100} shows that TS100 consists of 16 transistors in parallel, sharing a common source but with 16 different widths of gate G2 across cross-section A-A, leading to 16 individual drains.  The experimental setup involved applying $V_{\textrm{G1}} = -10$ V and $V_{\textrm{G2}} = V_{\textrm{ss}} = 0$ V.  The channel created by the applied gate voltages is essentially the same as a BC in a CCD, where electrons flow along $\phi_{\textrm{ch}}$ from the common source to each drain.  Measuring the onset of current flow gives $\phi_{\textrm{ch}}$.

A simulation of TS100 should be able to reproduce the measurements of $\phi_{\textrm{ch}}$ as a function of the width (W) of gate G2: the presence of G1 increasingly reduces $\phi_{\textrm{ch}}$ as the width of G2 across cross-section A-A reduces.  The range of W covers the range of pixel feature sizes in the Gaia CCDs: in the Gaia along (AL) scan (charge transfer) direction, the electrode widths are 2 and 3 $\mu$m but the 
potential below the gates is always formed by two adjacent gates (i.e. combined width of 5 $\mu$m due to the simultaneous clocking scheme); in the Gaia across (AC) scan direction, the anti-blooming drain (ABD) is 2 $\mu$m wide, the supplementary BC doping (SBC) widths are 3 and 4 $\mu$m, the ABD shielding is 4 $\mu$m wide and the BC doping is 21 $\mu$m wide\cite{seabroke2008,kohley2009}.  

The modelling input that electron distribution and $\phi$ are most sensitive to is doping.  e2v modelled their doping implant process using SSuprem3, the one-dimensional (1D) process simulation module in ATHENA, which is Silvaco's process simulation framework that ``enables process and integration engineers to develop and optimize semiconductor manufacturing processes".  This yielded a spread of simulated doping profiles:  net dopant density ($\rho$, donor ions cm$^{-3}$) as a function of depth ($z$, cm) into the device's Si layer.  The resulting spread was calibrated against Secondary Ion Mass Spectrometry (SIMS) measurements and empirically fit by a function of the following form:

\begin{equation}
\label{eq:density}
\rho = f \times 10^{az^{2} + bz + c} + \rho_{\textrm{ss}} \mbox{\hspace{1cm}where\hspace{1cm}} \rho_{\textrm{ss}} = \frac{1}{q \mu_{\textrm{p}} R},
\end{equation}

\noindent where $f$ is the dimensionless BC scale factor, $a$, $b$ and $c$ are constants proprietary to e2v, $\rho_{\textrm{ss}}$ is the substrate doping density (acceptor ions cm$^{-3}$), 
$q$ is electron charge (C), $\mu_{\textrm{p}}$ is hole mobility (480 cm$^{2}$V$^{-1}$s$^{-1}$) and $R$ is the substrate resistivity ($\Omega$cm).  

$R$ = 20 $\Omega$cm for TS100, which gives $\rho_{\textrm{ss}} = -6.51 \times 10^{14}$ acceptor ions cm$^{-3}$ (by convention donor and acceptor ions have positive and negative densities respectively because they are positively and negatively charged and thus can cancel out the effect of each other).  Note the Gaia CCDs use much higher resistivity Si: $R$ = 100 $\Omega$cm for the Astrometric Field (AF) and Blue Photometer (BP) CCDs and $R$ = 1500 $\Omega$cm for the Red Photometer (RP) and Radial Velocity Spectrograph (RVS) CCDs (A. Walker (e2v), private communication).  Nevertheless, like the Gaia AF and BP CCDs, TS100 also has the e2v standard depth of 16 $\mu$m, whereas the Gaia RP and RVS CCDs are e2v deep depletion devices with thicknesses of 40 $\mu$m.  Although the simulations presented in this paper are for a lower resistivity device (TS100) than the Gaia CCDs, the same methods presented here will be used to simulate the Gaia CCDs and so the benchmarking of these methods also applies to simulations of the Gaia CCD pixels. 

$f$ is a free parameter that cannot be easily physically measured.  Instead, it is varied in Equation \ref{eq:density} to produce a doping profile, which is substituted into Poisson's equation to solve for $\phi$.  The only part of the simulated internal $\phi$ distribution that can be compared to measurement is $\phi_{\textrm{ch}}$.  However, the difference between the measured and predicted $\phi_{\textrm{ch}}$ depends on flat-band voltage ($V_{\textrm{fb}}$), which cannot be independently measured.  Therefore, models need to predict a parameter that is independent of $V_{\textrm{fb}}$ and can be measured: $\phi_{\textrm{chp}}$.  Using e2v's in-house simulation software, written by M. Robbins\cite{Robbins1992}, $f$ is varied to satisfy $\phi_{\textrm{chp}}$ = 4.6 V, which is the standard value measured by e2v.  This is assumed to apply to TS100 because $\phi_{\textrm{chp}}$ is not strongly dependent on $R$.  The resulting value of $f$ is used in Equation \ref{eq:density} to derive a doping profile applicable to TS100, which is read into ATLAS, Silvaco's device simulation framework: ``ATLAS is a physically based two and three dimensional device simulator. It predicts the electrical behaviour of specified semiconductor structures and provides insight into the internal physical mechanisms associated with device operation".  The primary user-inputs to ATLAS are device geometry, materials and the doping profile(s).

\begin{figure}[htbp]
\begin{center}
 \includegraphics[width=0.9\textwidth]{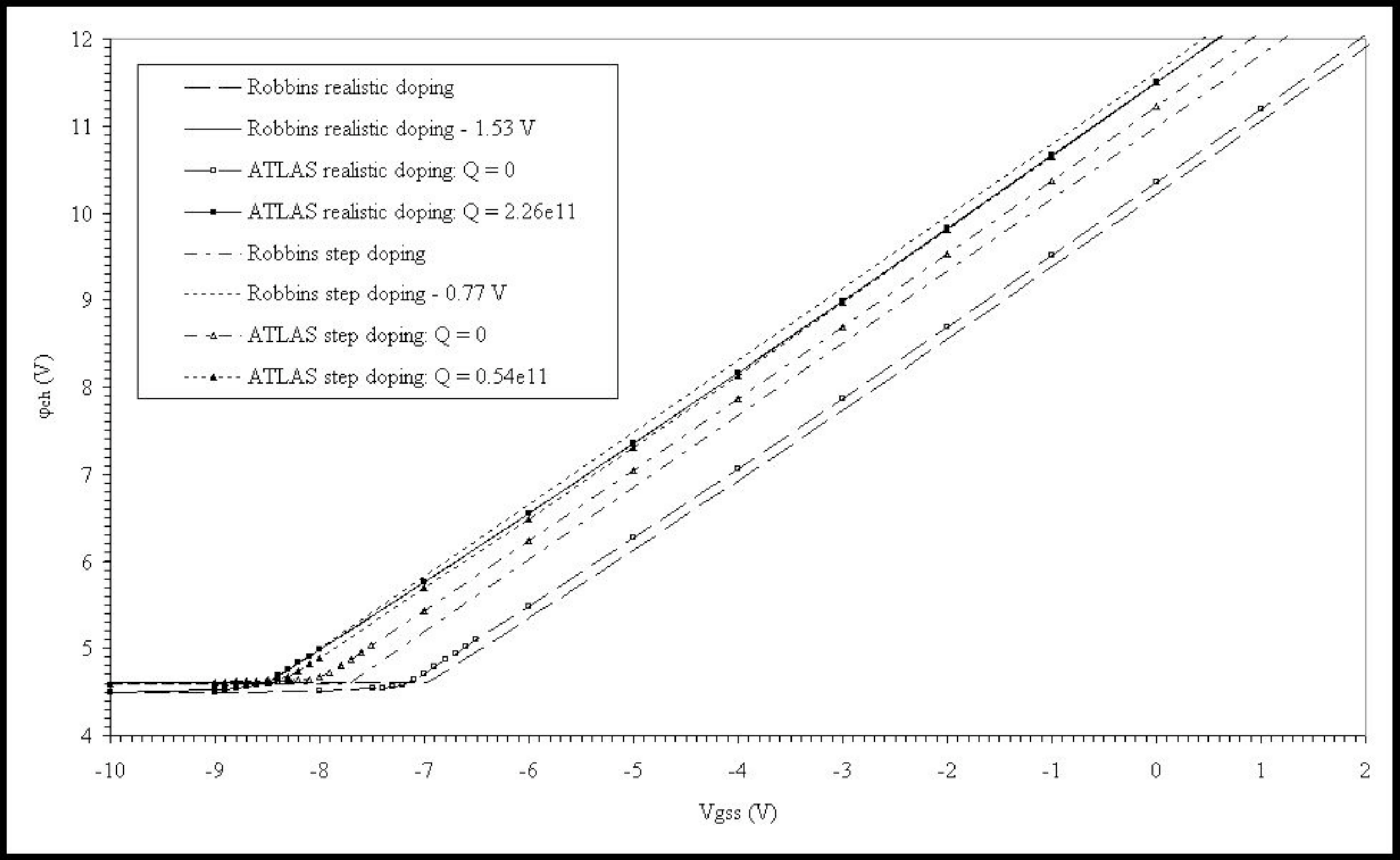}
 \includegraphics[width=0.9\textwidth]{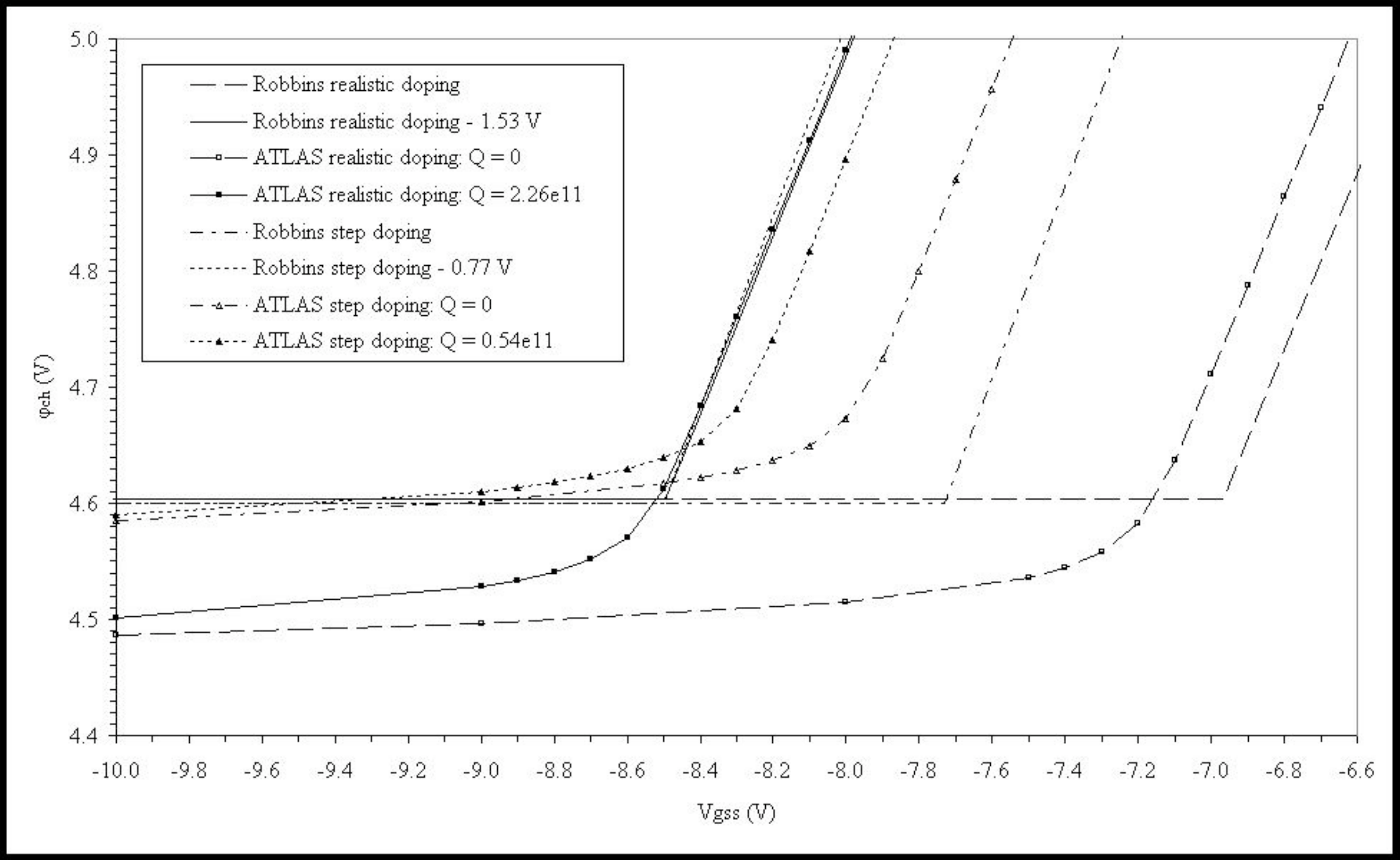}
 \end{center}
\caption{Plots of Robbins 1D and ATLAS planar test transistor models simulating $\phi_{\textrm{ch}}$ against $V_{\textrm{gss}}$, showing $\phi_{\textrm{ch0}}$, $\phi_{\textrm{chp}}$ and $V_{\textrm{gssp}}$ ({\it top}) and zoomed in on the $\phi_{\textrm{chp}}$ and $V_{\textrm{gssp}}$ region ({\it bottom}).}
\label{fig:q}
\end{figure}

Robbins' model implements the simplest possible numerical method to solve Poisson's equation in 1D so it does not attempt to physically model $V_{\textrm{fb}}$.  To derive $V_{\textrm{fb}}$ requires modelling the test transistor that e2v used to measure $V_{\textrm{gssp}}$.  Robbins models this in 1D by applying voltage to a layer of SiO$_{2}$ on top of a slab of doped Si.  Figure \ref{fig:q} shows that the Robbins model pins at $V_{\textrm{gssp}} = -6.97$ V, whereas e2v measured $V_{\textrm{gssp}} \approx -8.5$ V.  This value was not measured from TS100 itself but is an average value of the batch, which is consistent with the batch-to-batch average of $-8.1$ V for $R$ = 20 $\Omega$cm.  There is a spread of channel parameters wafer-to-wafer and batch-to-batch so the actual value that TS100 actually pins at is not known.  Nevertheless, this is the best information available with which to calibrate the Robbins model in order to approximately model TS100.   If ``Robbins realistic doping $-$ 1.53 V" is plotted in Fig.~\ref{fig:q} to represent a model that does include $V_{\textrm{fb}}$, it shows that the Robbins model gives the same $\phi$ as the Robbins realistic doping line when $V_{\textrm{gss}} - V_{\textrm{fb}}$ is applied to the gate, where $V_{\textrm{fb}} = -1.53$ V (by convention $V_{\textrm{fb}}$ is a negative number).  ATLAS physically models $V_{\textrm{fb}}$ 
via the free parameter fixed oxide charge ($Q$), which is defined at all insulator-semiconductor interfaces per unit area (C cm$^{-2}$).  

The same Robbins test transistor model doping profile and oxide thickness was used in the corresponding ATLAS model, which has to be two-dimensional (2D).  Therefore the main dimension is depth with an arbitrary distance in an orthogonal device dimension.  The structure does not change in this direction and so the ATLAS simulation is planar.  Figure \ref{fig:q} shows the models exhibit similar Non-Inverted Mode Operation (NIMO, $V_{\textrm{gss}} > V_{\textrm{gssp}}$) behaviour: the ``ATLAS realistic doping: $Q = 0$" line has the same NIMO gradient as the Robbins model, only offset by $\approx$0.2 V.  It shows that $Q$ needs to be calibrated to the ``Robbins realistic doping $-$ 1.53 V" line in order for ATLAS to model $V_{\textrm{fb}} = -1.53$ V.  

The bottom plot of Fig.~\ref{fig:q} highlights a difference in behaviour between the Robbins model and ATLAS in Inverted Mode Operation (IMO, $V_{\textrm{gss}} < V_{\textrm{gssp}}$).  The Robbins model produces straight IMO lines because it is an electrostatic model, simulating $\phi$ differences due to space charge regions.  ATLAS does not produce straight IMO lines because it is a full semiconductor model, simulating electron and hole carrier concentrations.  The second order effect of more and more holes flowing to the Si-SiO$_{2}$ interface causes the ATLAS IMO lines to curve.  This means it is more difficult to define $\phi_{\textrm{chp}}$ and $V_{\textrm{gssp}}$ from ATLAS simulations.  Because of this and the fact that the TS100 G2 gates (and Gaia CCDs) are run in NIMO, $Q$ is varied to align the NIMO parts of the ATLAS line and the ``Robbins realistic doping $-$ 1.53 V" line.  $Q = 2.26 \times 10^{11}$ C cm$^{-2}$ brings them into excellent agreement.  However, the models cannot be compared in an absolute sense to the TS100 measurements because TS100's $V_{\textrm{gssp}}$ is not known, which means that TS100's $V_{\textrm{fb}}$ cannot be derived for the Robbins model nor TS100's $Q$ for ATLAS.  Nevertheless, exactly calibrating $Q$ to the Robbins model has the advantage of allowing a direct 2D comparison of ATLAS and the Robbins model, while only being able to compare to TS100 measurements in a relative sense.

The average doping profile in Equation \ref{eq:density} only applies to the BC part of the pixel and not profiles through SBCs and ABDs, which are features in the Gaia image pixels.  To avoid doping and $\phi$ discontinuities in the AC pixel dimension, ideally ATLAS  would use a 2D doping file derived from 2D ATHENA simulations as input (doping does not change in the AL direction and thus 3D doping would be superfluous).  However, e2v have not modelled the implant processes for the BC, SBC and ABD simultaneously  in SSuprem4, ATHENA's 2D process simulation software.  In the absence of such a 2D doping file, the easiest and quickest way to simulate complex doping profiles is using step profile doping.  Realistic doping profiles can be approximated by uniformly doped boxes: e.g. the SBC part of the pixel would have one box for the SBC itself and one for the substrate (the constant density of which is given by Equation \ref{eq:density}), which is less than the SBC box doping forming a step profile.  However, a BC realistic doping profile adjacent to a SBC step doping profile would cause doping and $\phi$ discontinuities.  Thus, the whole Gaia image pixel needs to be modelled with step doping.  Both realistic and step doping ATLAS models are benchmarked in this paper to highlight the effect of different doping approximations.

The TS100 BC box parameters are the BC doping (n-type, implant donor dopant density $N_{\textrm{D}}$, cm$^{-3}$) and the BC implant depth ($d$, cm), over which $N_{\textrm{D}}$ is uniformly distributed.  ATHENA modelling by K. Ball (e2v) finds $d$ = 0.54 $\mu$m.  Like $f$ in the realistic doping case, the value of $N_{\textrm{D}}$ needs to satisfy $\phi_{\textrm{chp}}$ = 4.6 V.  $N_{\textrm{D}}$ is derived using the step doping model, which, assuming full depletion, allows the $\phi$ distribution within the gate insulator and Si region to be analytically obtained\cite{janesick2001}.  Using e2v's in-house simulation software implementation of the step doping model (written by M. Robbins), $N_{\textrm{D}} = 2.76 \times 10^{16}$ cm$^{-3}$.  The right plot of Fig.~\ref{fig:pot_profile} compares the resulting step doping profile with the realistic one.  The BC box in the step doping model should have the same number of donor dopant ions as the realistic doping profile.  Integrating the BC part of Equation \ref{eq:density} (by excluding the $\rho_{\textrm{ss}}$ term) with respect to $z$ calculates the concentration of dopant ions in the BC per unit area of the device in the plane perpendicular to $z$ ($\rho_{\textrm{BC}}$, cm$^{-2}$).  $N_{\textrm{D}} = \rho_{\textrm{BC}}/d = 2.65 \times 10^{16}$ cm$^{-3}$, which is consistent with the uniformly doped BC box value. 


Figure \ref{fig:q} shows the step doping models also exhibit similar NIMO behaviour: the ``ATLAS step doping: $Q = 0$" line has the same NIMO gradient as the Robbins step doping model.  The different methods the two models employ, the Robbins step doping model is a 1D analytical solution of Poisson's equation while the ATLAS planar step doping model is a 2D numerical solution of Poisson's equation, should produce the same results.  The offset of $\approx$0.3 V shows fairly good agreement.  With no $V_{\textrm{fb}}$ or $Q$, $V_{\textrm{gssp}}$(step)  $< V_{\textrm{gssp}}$(realistic).  Therefore the $V_{\textrm{fb}}$ required to shift the Robbins step doping line to pin at $-8.5$ V is less than the realistic doping value of $V_{\textrm{fb}}$.  The  resulting ``Robbins step doping $-$~0.77~V" line has a slightly steeper gradient than the realistic doping line.  Because the different doping models are diverging around $\phi_{\textrm{ch0}}$, the ATLAS step doping model has its $Q$ calibrated against the ``Robbins realistic doping $-$ 1.53 V" to make the step doping model as similar as possible to the realistic doping model around $\phi_{\textrm{ch0}}$, which gives $Q = 0.54 \times 10^{11}$ C cm$^{-2}$.  However, Fig.~\ref{fig:q} shows that the excellent agreement between the ATLAS step and realistic doping models around $\phi_{\textrm{ch0}}$ is at the expense of good agreement around $\phi_{\textrm{chp}}$.  The bottom plot of  Fig.~\ref{fig:q} shows that the IMO $\phi_{\textrm{ch}}$ of the ATLAS step doping model agrees with the Robbins models at $\approx$4.6 V, whereas the realistic doping models are offset by $-0.1$ V at $\approx$4.5 V.

\begin{figure}[htbp]
\begin{center}
 \includegraphics[width=0.9\textwidth]{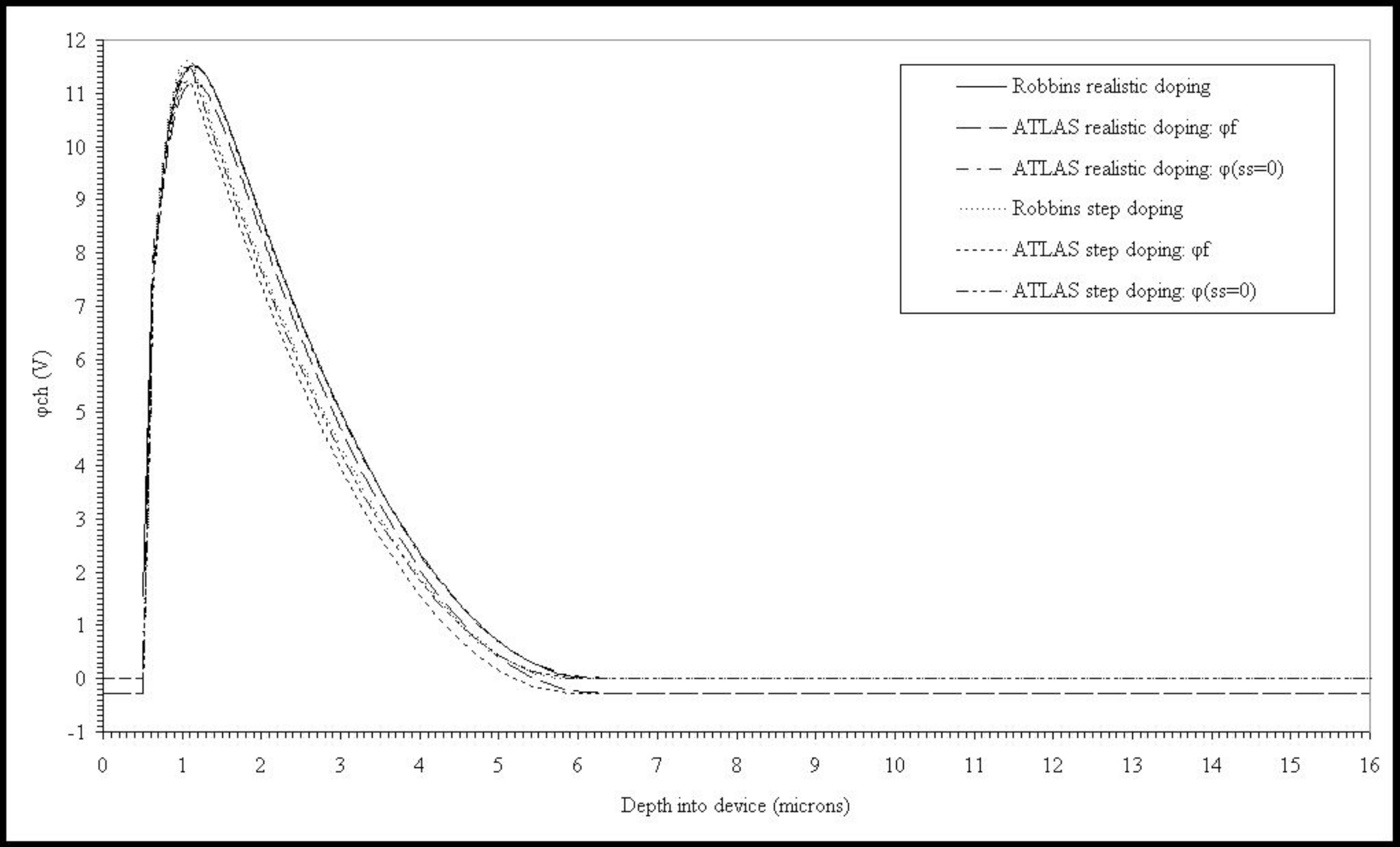}
 \includegraphics[width=0.9\textwidth]{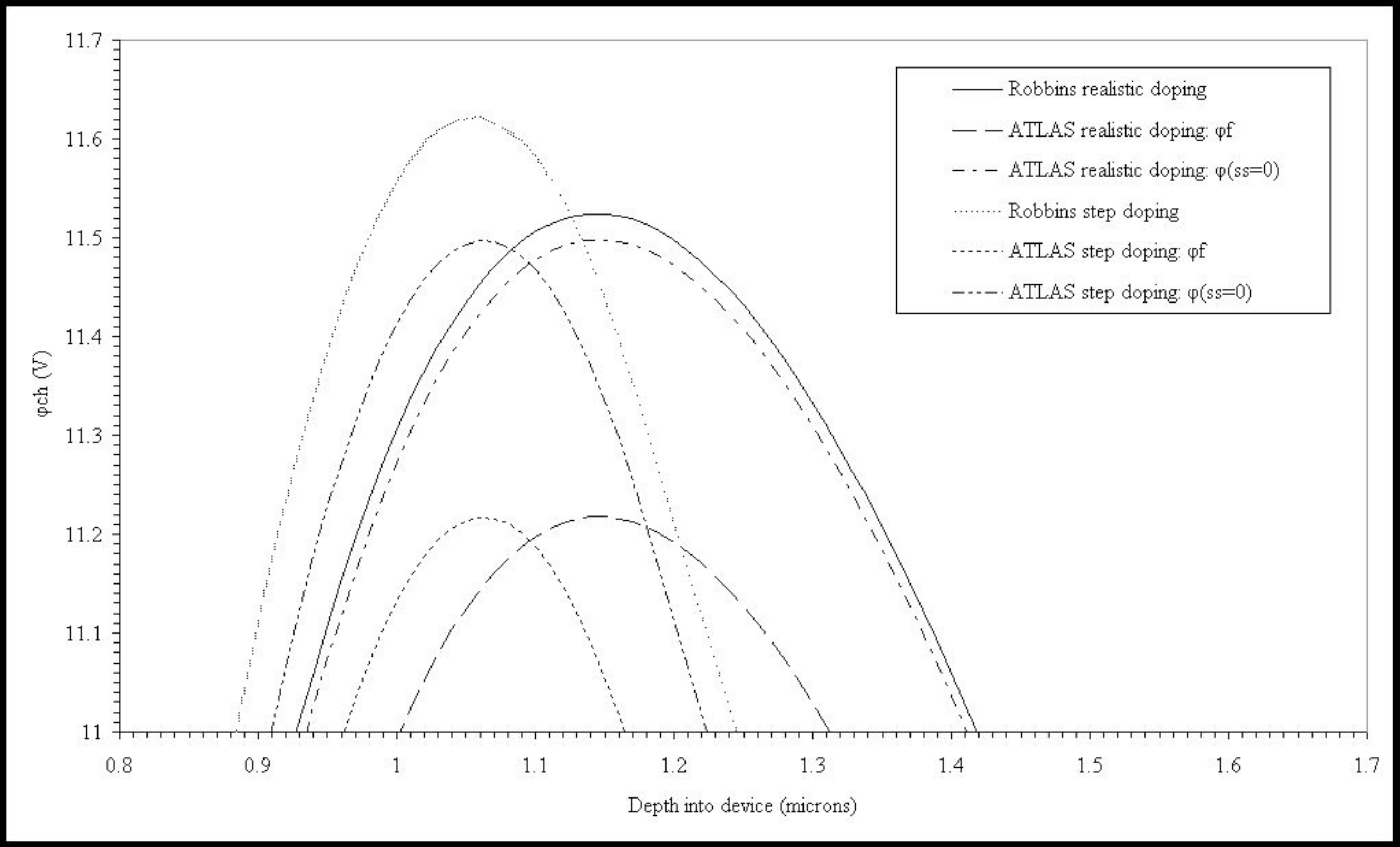}
 \end{center}
\caption{Plots of Robbins 1D and ATLAS planar test transistor models simulating NIMO $\phi$ against depth into device, comparing the $\phi$ resulting from realistic and step doping profiles ({\it top}) and zoomed in on the $\phi_{\textrm{ch0}}$ region ({\it bottom}).  The top plot shows the full depth of TS100: electrodes (depth 0-0.5 $\mu$m), SiO$_{2}$ (depth 0.5-0.63 $\mu$m) and Si (depth 0.63-16 $\mu$m).  The Si-SiO$_{2}$ interface is 0.63 $\mu$m deep.}
\label{fig:profiles}
\end{figure}

``The reference $\phi$ can be defined in various ways. For ATLAS, this is always the intrinsic Fermi $\phi$ ($\phi_{\textrm{i}}$)" (ATLAS UserÕs Manual).  In an intrinsic semiconductor, $\phi_{\textrm{i}} = \phi_{\textrm{f}}$, where $\phi_{\textrm{f}}$ is the Fermi $\phi$, which is the electrochemical $\phi$ equivalent to voltage.  However, in extrinsic (doped) semiconductors, referencing $\phi_{\textrm{f}}$ to $\phi_{\textrm{i}}$ means that $\phi_{\textrm{i}}$ = 0 V in the following equation, which only applies in a p-type substrate:

\begin{equation}
\label{eq:ptype}	
\phi_{\textrm{f}} = \phi_{\textrm{i}} + \frac{kT}{q} \textrm{ln}\left(\frac{N_{\textrm{A}}}{n_{\textrm{i}}}\right), 
\end{equation}

\noindent where $k$ is Boltzmann's constant (J~K$^{-1}$), $T$ is temperature (it is assumed that TS100 measurements were conducted at room temperature: 293 K), $N_{\textrm{A}}$ is acceptor doping density ($N_{\textrm{A}} = |\rho_{\textrm{ss}}|$) and $n_{\textrm{i}}$ is the intrinsic carrier density ($\approx$7~$\times 10^{9}$ cm$^{-3}$).  Substituting the values for TS100 into Equation \ref{eq:ptype} gives $\phi_{\textrm{f}} = -0.28$ V, which is the same value seen in the substrate of the $\phi_{\textrm{f}}$ profiles in Fig.~\ref{fig:profiles}.  With $\phi_{\textrm{ss}} = -0.28$ V, for $V_{\textrm{gss}}$ to equal zero, $V_{\textrm{g}}$ must also equal $-0.28$ V.  The Robbins model profiles have their $\phi$ referenced to $\phi_{\textrm{ss}} = 0$.  Therefore, to directly compare both model outputs, the ATLAS simulations were also referenced to $\phi_{\textrm{ss}} = 0$ by adding 0.28 V to every mesh point through the gate, oxide and Si of the ATLAS profiles and plotted in Fig.~\ref{fig:profiles} (this was also done for $V_{\textrm{gss}}$ and $\phi_{\textrm{ch}}$ for every ATLAS simulation in Fig.~\ref{fig:q}).  

The bottom plot of Fig.~\ref{fig:profiles} shows that the $\phi$ distribution around $\phi_{\textrm{ch}}$ differs between step and realistic doping.  Both the step doping models produce a shallower and narrower peak in terms of depth into the Si.  The Robbins step doping model produces a higher $\phi_{\textrm{ch}}$ than both the realistic doping models.  Because the step doping $\phi$ distribution around $\phi_{\textrm{ch}}$ is constant in shape and depth and is scaled in $\phi$ by $Q$, it is possible to remove one of the differences between the step and realistic doping $\phi$ distributions by calibrating $Q$ to make the ATLAS step doping model $\phi_{\textrm{ch}}$ agree with the $\phi_{\textrm{ch}}$ of the realistic doping models.  The other differences in the $\phi$ distributions around $\phi_{\textrm{ch}}$ are intrinsic to the differences between the realistic and step doping profiles. 

\begin{figure}[htbp]
\begin{center}
 \includegraphics[width=0.49\textwidth]{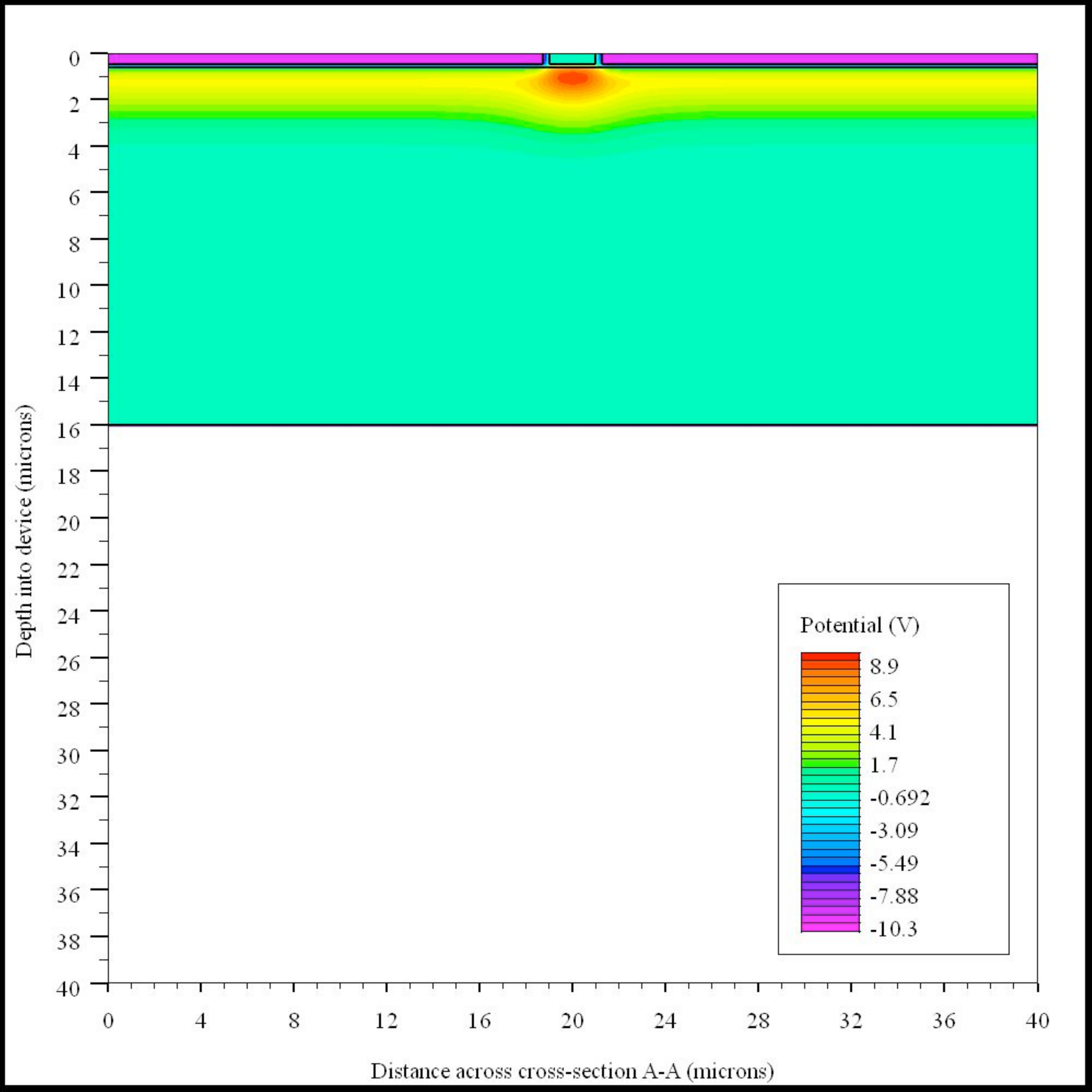}
 \includegraphics[width=0.49\textwidth]{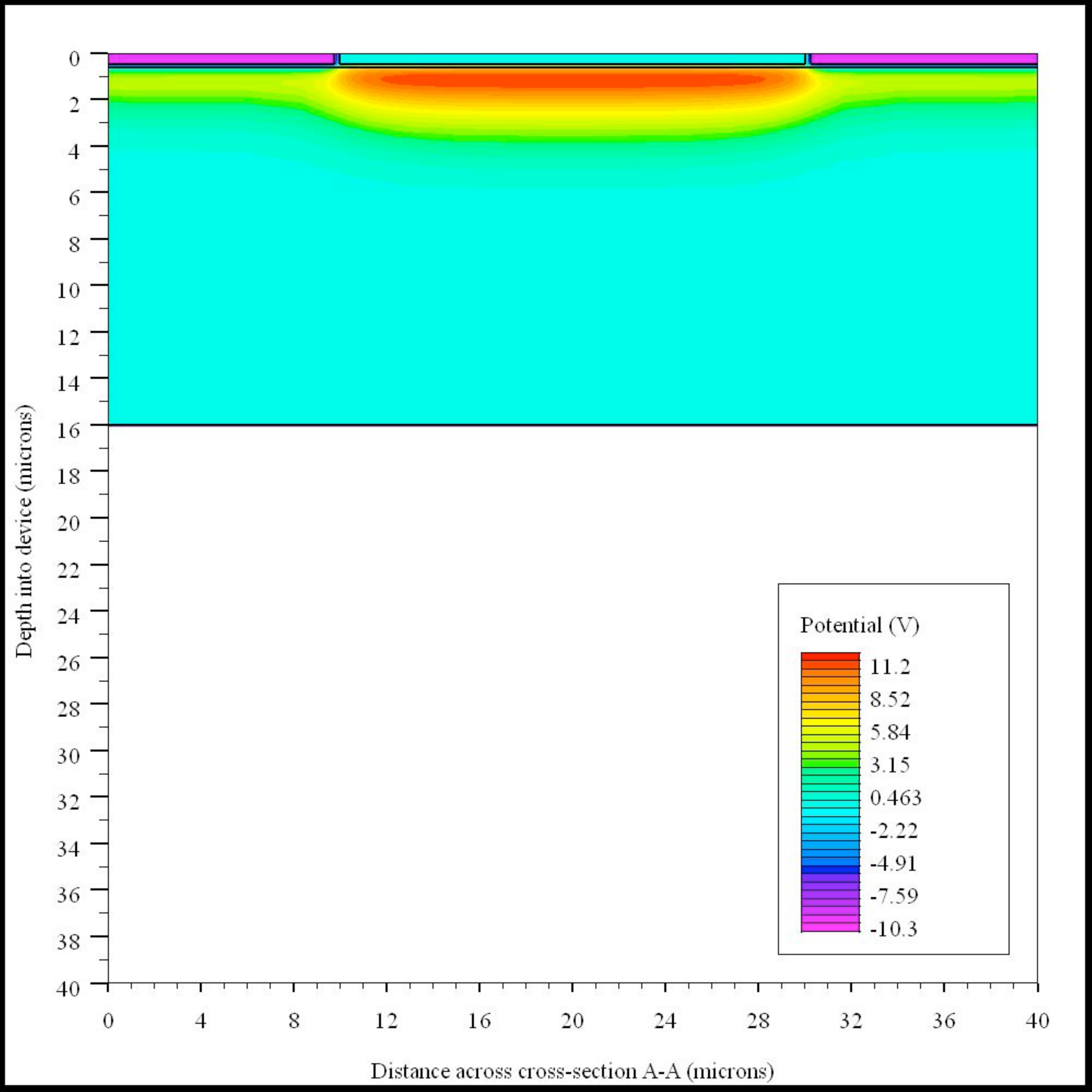}
 \end{center}
\caption{Silvaco TonyPlots of the 2D ATLAS model showing the two ends of TS100: middle gate G2 with a width of 2~$\mu$m ({\it left}) and 20 $\mu$m ({\it right}) with $\phi_{\textrm{f}}$ contours overlaying the electrodes, SiO$_{2}$ layer and the Si, due to $V_{\textrm{G1}} = -10.28$ V either side of $V_{\textrm{G2}} = -0.28$ V.  The gaps between the electrodes are 0.25 $\mu$m and filled with oxide.}
\label{fig:w}
\end{figure}

\begin{figure}[htbp]
\begin{center}
 \includegraphics[width=0.9\textwidth]{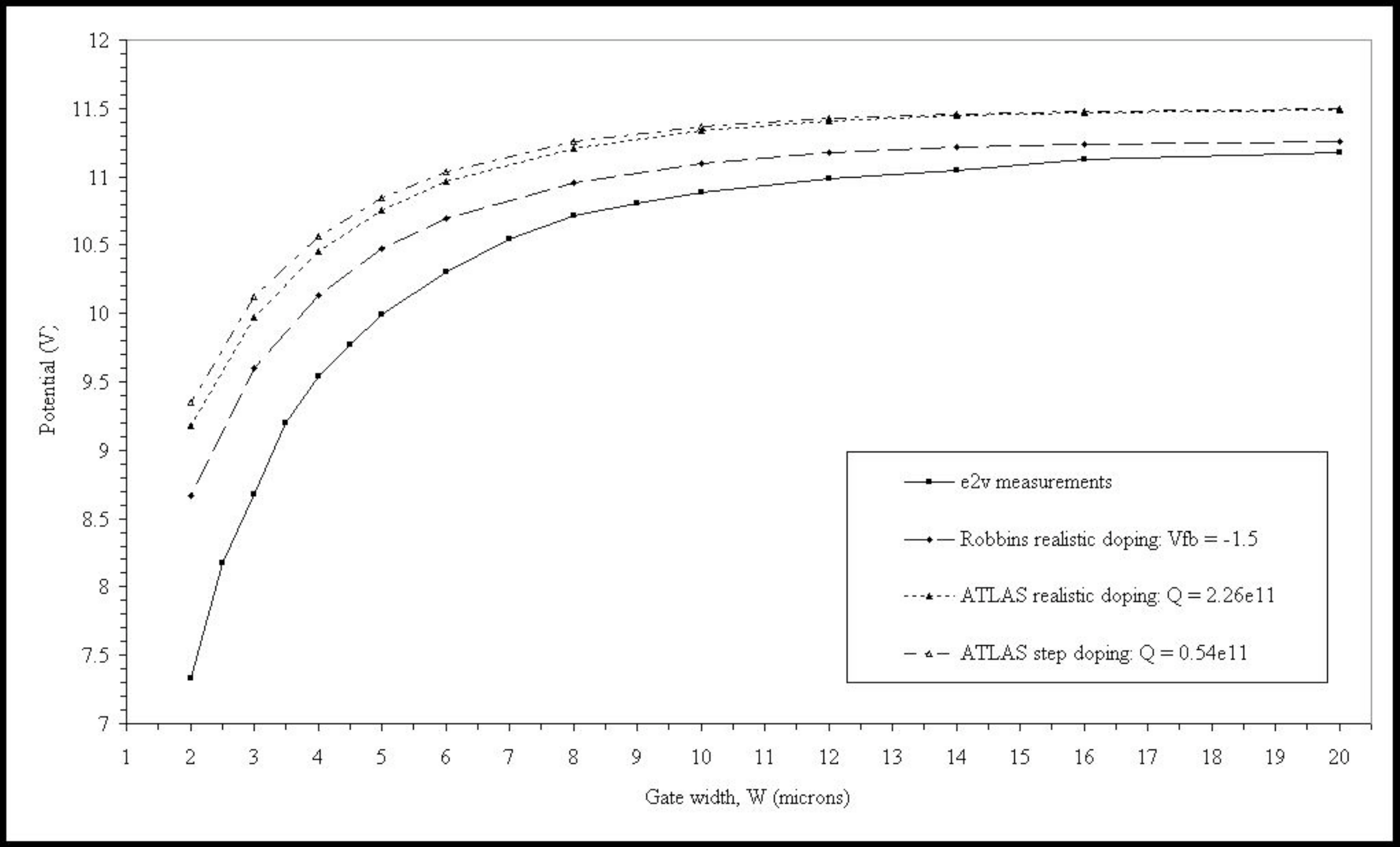}
 \includegraphics[width=0.9\textwidth]{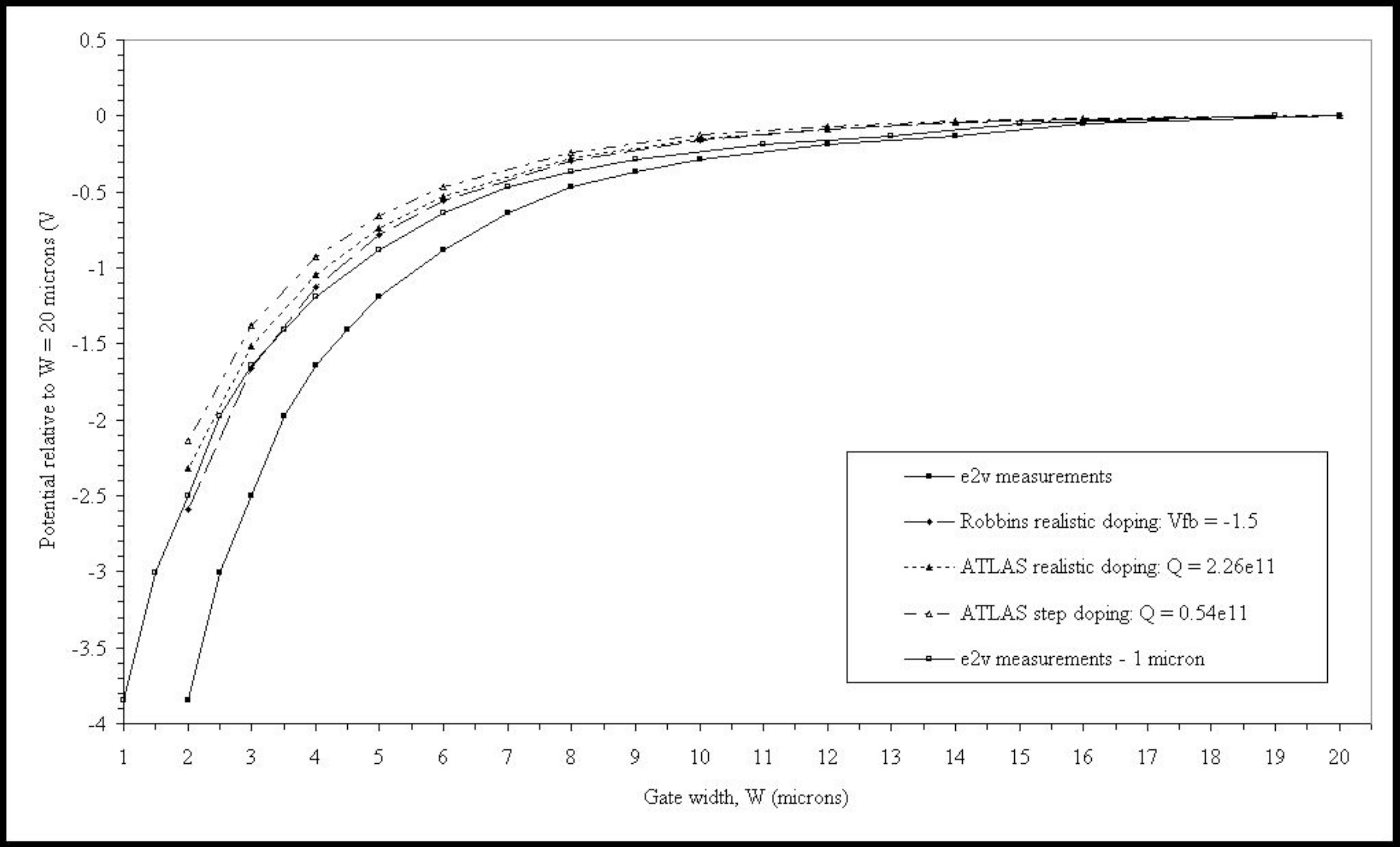}
 \end{center}
\caption{Plots of $\phi_{\textrm{ch}}$ as a function of G2 gate width, comparing measurements of TS100 by e2v to 2D simulations of TS100 by the Robbins and ATLAS models, absolutely ({\it top}) and relatively ({\it bottom}).}
\label{fig:abs_rel}
\end{figure}

Having derived the realistic and step doping profiles and $V_{\textrm{fb}}$ from 1D Robbins simulations and $Q$ from planar ATLAS simulations, these values can now be applied in 2D models of TS100.  TS100 only needs to be modelled in 2D because the length of gate G1 in Fig.~\ref{fig:ts100} in the charge flow direction is long enough for 3D effects to be negligible.  Both the 2D Robbins model\cite{robbins2001} and ATLAS model simulate TS100 by modelling the cross-section A-A in Fig.~\ref{fig:ts100} but only including one G2 gate width at a time.  The cross-section A-A shows that G1 overlaps G2.  This does not need to be modelled because only conductors in contact with the insulator over the Si determine the $\phi$ distributions in the Si.  Therefore, Fig.~\ref{fig:w} shows how the gates were simulated and how the $\phi$ under the G1 gates pinches off the $\phi$ under G2 when G2 is narrow (left plot) compared to when it is wide (right plot).  $\phi_{\textrm{ch}}$ is measured from a $\phi$ profile through the centre of each G2 gate width and is plotted in Fig.~\ref{fig:abs_rel}.  In ATLAS, the maximum $\phi$ under G2 is actually $\phi_{\textrm{ch}} - \phi_{\textrm{ss}}$, where $\phi_{\textrm{ss}} = -0.28$ V.  

The models simulate $\phi$(internal), which cannot be directly measured.  $\phi$(transistor measured) is less than $\phi$(internal) because it does not take into account the ``diode drop".  This is the $\phi$ required to surmount the p-n junction barrier (diode) and commence conduction, called built-in $\phi$ ($\phi_{\textrm{bi}}$).    Therefore, $\phi$(measured) does not measure the diode drop because it has ``used up''  $\phi_{\textrm{bi}}$ to allow the measurement to be made and so $\phi_{\textrm{bi}}$ needs to be added back on to the $\phi$(measured) to make it up to the value of $\phi$(internal), to compare directly with the models.    The exact value of $\phi_{\textrm{bi}}$ depends on the doping profile of the transistor's source and drain but the standard value of 0.60~V is assumed.  Thus all the e2v measurements in Fig.~\ref{fig:abs_rel} have had 0.60~V added to them.  

Figure \ref{fig:abs_rel} shows all the models and measurements have $\phi_{\textrm{ch}}$ values that plateau at large G2 gate widths.  The top plot of Fig.~\ref{fig:abs_rel} shows that at W = 20 $\mu$m, the ATLAS realistic and step doping 2D models give $\phi_{\textrm{ch}}= 11.50$~V.  This is the same as the $\phi_{\textrm{ch0}}$ = 11.50 V produced by the planar ATLAS (and 1D Robbins) models in Figs.~\ref{fig:q} and \ref{fig:profiles}.  However, one may have expected a difference in $\phi_{\textrm{ch}}$ between the ATLAS planar and 2D simulations because the planar simulation does not include any fringing fields (because there is only one gate), whereas the 2D simulation does include fringing fields, which can decrease $\phi_{\textrm{ch}}$.  This decrease is seen comparing the Robbins 1D and 2D models: the top plot of Fig.~\ref{fig:abs_rel} gives $\phi_{\textrm{ch}} = 11.26$~V at W = 20 $\mu$m, which is 0.24 V less than the $\phi_{\textrm{ch0}}$ = 11.50 V produced by the Robbins 1D model.  The difference between the ATLAS and Robbins 2D models could be related to how the models simulate $\phi$ in the oxide gaps between the electrodes: ATLAS solves Poisson's equation at every mesh point, whereas the Robbins model linearly interpolates between the electrodes.  The ATLAS realistic doping model produces $\approx$0.1 V less than the Robbins models in 1D at $V_{\textrm{gss}} = -10$ V in Fig.~\ref{fig:q}, suggesting in 2D one may have expected the Robbins model to produce higher $\phi$ values than ATLAS, rather than vice versa as seen in Fig.~\ref{fig:abs_rel}.

The ATLAS step doping model, which was deliberately calibrated to the ATLAS planar realistic doping model, shows excellent agreement with the ATLAS realistic doping model in 2D at large gate widths but diverges as W decreases (and fringing fields become closer in size to the gate width) until reaching its maximum offset of $\approx$0.2~V at W = 2 $\mu$m.  This suggests that ATLAS models using step doping profiles do not simulate fringing fields as well as models using realistic doping profiles.  Gaia charge packets will sit in $\phi$ defined by two adjacent gates with combined width of 5 $\mu$m wide so modelling the Gaia pixel with step doping could introduce a $\phi$ offset of 0.1 V but this is much less than the expected variation of up to $\pm$15\% from Gaia CCD production spreads\cite{burt2003}. 

The absolute differences between the models are not as important as the relative differences because electrons (and holes) only ``see" relative changes in $\phi$ rather than absolute values.  The bottom plot of Fig.~\ref{fig:abs_rel} shows there is excellent agreement between all the models at large gate widths when each model is normalised to its $\phi_{\textrm{ch}}$ when W = 20 $\mu$m.  The Robbins and ATLAS models diverge at smaller gate widths to a maximum offset of $\approx$0.3~V at W = 2 $\mu$m. Again, this divergence is indicative of the differences of how the models simulate the oxide gaps and the 2D effect of fringing fields.  

The gate widths plotted in Fig.~\ref{fig:abs_rel} are the design values.  However, manufacturing tolerances mean that all dimensions of the electrodes can be larger or, more likely, smaller than the nominal value by up to 1 $\mu$m due to under- and over-etching respectively.  All gates will differ from their nominal dimensions by the same amount.   Plotting the e2v measurements minus 1 $\mu$m in the bottom plot of Fig.~\ref{fig:abs_rel} brings the measurements into much better agreement with both the Robbins and ATLAS realistic doping models than previously and provides a plausible explanation for the relative $\phi$ offset between the measurements and the models.





\section{CONCLUSIONS AND FUTURE WORK}
\label{sec:conc}  

The aim of this series of papers is to derive the electron distribution within a charge packet, as a function of 3D position within the different Gaia CCD pixels and as a function of the number of electrons in the charge packet, to feed into Gaia radiation damage correction models.  This paper has presented the first results of using the Silvaco ATLAS device simulation software, benchmarking it against other simulation software (Robbins' models) and physical measurements of a test structure.  

Inputting the correct doping profile, pixel geometry and materials into ATLAS and comparing to the corresponding 1D Robbins model reveals that ATLAS has a free parameter, fixed oxide charge ($Q$) that needs to be calibrated.  Because the measurement required to calibrate 1D models to e2v's test structure TS100 ($V_{\textrm{gssp}}$) was not made, $Q$ was calibrated against the Robbins 1D model.  This was done in the Non-Inverted Mode Operation (NIMO) part of the simulated $\phi_{\textrm{ch}}$-$V_{\textrm{gss}}$ curve because of the different behaviour of the models in the Inverted Mode Operation (IMO) part and the fact that the TS100 G2 gate and Gaia CCDs operate in NIMO.

Comparison of ATLAS and Robbins model potential ($\phi$) profiles highlights the different reference potentials ($\phi$) of the models.  ATLAS is referenced to the intrinsic Fermi $\phi$, while the Robbins model is referenced to $\phi_{\textrm{ss}} = 0$.  For TS100 with $V_{\textrm{gss}}$(G1) = $-10$ V, $V_{\textrm{gss}}$(G2) = 0 V and $V_{\textrm{ss}} = 0$ V, this meant the ATLAS model applied $V_{\textrm{g}}$(G1) = $-10.28$ V and $V_{\textrm{g}}$(G2) = $-0.28$ V because $\phi_{\textrm{ss}} = -0.28$ V even with $V_{\textrm{ss}} = 0$ V applied.  Hence to simulate the Gaia $\phi$ distribution under $V_{\textrm{g}}$(high) = 10 V and $V_{\textrm{g}}$(low) = 0 V ($V_{\textrm{ss}}$ = 0 V) will require applying $V_{\textrm{g}} + \phi_{\textrm{ss}}$ to each gate.  This will mean that electrons in the Gaia 3D pixel ATLAS simulations will sit in slightly different absolute $\phi$ values compared to reality.  The electron distribution will not differ from reality as long as the relative $\phi$ values between adjacent electrodes are consistent.

Because ATLAS was calibrated to the Robbins model in 1D, this permitted a direct absolute comparison of the models in 2D.  This reveals a small $\phi_{\textrm{ch}}$ difference between the models in 2D, ranging from 0.24-0.5 V going from the widest to the narrowest gate.   As the $V_{\textrm{fb}}$ and $Q$ used in the 2D models are not specific to TS100, the Robbins and ATLAS models could only be compared to measurements in a relative sense.  While the models differ slightly in their predicted  $\phi_{\textrm{ch}}$ under narrow gate widths, the general trend of both models follows the physical measurements of TS100 well (after correction for a suspected systematic offset in TS100 itself).  

It was found that step doping produces different $\phi$ distributions around $\phi_{\textrm{ch}}$ to realistic doping.  As the realistic doping is more accurate than step doping, the $\phi_{\textrm{ch}}$ produced by step doping will not be as physically realistic as if realistic doping had been used.  Consequently, the electron distributions produced by step doping may not be as physically realistic as if realistic doping had been used.  However, it is not yet clear to what extent changes in $\phi$ affect electron distributions.  This is will be investigated in the course of this work.  In the absence of realistic doping profiles for Gaia image pixels, these pixels will be modelled with step doping profiles.  Although, this will provide only an approximation of physically-derived electron distributions, it will be an improvement on arbitrarily-chosen electron distribution functions used in previous Gaia radiation damage models\cite{seabroke2008}.

Therefore, ATLAS has been successfully benchmarked against a device with lower resistivity (TS100) than the Gaia CCDs.  Because the simulation methods are the same, regardless of resistivity, the benchmarking results are equally applicable to Gaia simulations, identifying how to use ATLAS to model Gaia pixels and the effect of using step rather than realistic doping.  This is the first attempt by anyone to model the Gaia pixel to derive its electron distributions.  It may also be the first time this has been done for any e2v pixel.  If this is not the first attempt by anyone to model an e2v pixel, it is the most complex e2v pixel to be modelled, as the Gaia image pixel is the most complex pixel architecture ever made by e2v, because it includes so many different features: buried channel, supplementary buried channel and anti-blooming drain and shield.  A realistic microscopic Gaia pixel model will not only provide inputs to Gaia radiation damage correction models but be able to simulate and physically explain the results found by detailed measurements of close-reject Gaia Flight Model CCDs\cite{kohley2009}.

\acknowledgments     
 
The Silvaco ATLAS software license fee is generously funded by the UK VEGA Gaia Data Flow System grant thanks to F. van Leeuwen and G. Gilmore (Institute of Astronomy, Cambridge).  Thank you very much to A. Walker and K. Ball (e2v) for kindly providing e2v's experimental measurements of channel parameters, D. Green and A. Nejm (Silvaco Europe Ltd, Cambridge) for providing Silvaco software support, M. Davidson (Edinburgh) for summarizing the Gaia PSF and LSF calibration work and R. Kohley (ESAC) for clarifying the text.


\bibliography{seabroke_spie2009}   
\bibliographystyle{spiebib}   

\end{document}